\documentclass[twocolumn,amsmath,nofootinbib,amssymb]{revtex4}
\usepackage{epsfig}
\usepackage{graphicx}% Include figure files
\usepackage{dcolumn}% Align table columns on decimal point
\usepackage{bm}% bold math
\usepackage{color}

\newcommand{\lsim}{\mbox{\raisebox{-.9ex}{~$\stackrel{\mbox{$<$}}{\sim}$~}}}
\newcommand{\gsim}{\mbox{\raisebox{-.9ex}{~$\stackrel{\mbox{$>$}}{\sim}$~}}}

\def\thebiblio#1{
\begin{center}\bf \large References
\end{center}
\list
{[\arabic{enumi}]}{\settowidth\labelwidth{#1.}\leftmargin\labelwidth
 \advance\leftmargin\labelsep
 \usecounter{enumi}}
 \def\newblock{\hskip .11em plus .33em minus -.07em}
 \sloppy
 \sfcode`\.=1000\relax}

\begin{document}
\preprint{}
\title{%
Can a vector field be responsible for the curvature perturbation 
in the Universe?
}
% Force line breaks with \\

\author{Konstantinos Dimopoulos}%
\email{k.dimopoulos1@lancaster.ac.uk}
\affiliation{%
Physics Department, Lancaster University\\}%

\date{\today}% It is always \today, today,
             % but any date may be explicitly specified

\begin{abstract}
I investigate the possibility that the observed curvature perturbation is due
to a massive vector field. To avoid generating a large scale anisotropy the
vector field is not taken to be driving inflation. Instead it is
assumed to become important after inflation when it may dominate the Universe
and imprint its perturbation spectrum before its decay, as in the curvaton 
scenario. It is found that, to generate a scale invariant spectrum of 
perturbations, the mass-squared of the vector field has to be negative and 
comparable to the Hubble scale during inflation. After inflation 
the mass-squared must become positive so that the vector field engages into 
oscillations. It is shown that, such an oscillating vector field  behaves as 
pressureless matter and does not lead to large scale anisotropy when
it dominates the Universe. The possibility of realising this scenario in 
supergravity is also outlined.
\end{abstract}

\maketitle

\section{Introduction}
Observations of the curvature perturbation in the Universe strongly suggest
that it is generated during inflation by the gravitational production of 
particles. The field, whose quantum fluctuations are responsible for the
particle production is typically considered to be a scalar field; one of the
many flat directions that are envisaged in theories beyond the standard model.

Very little has ever been discussed about gravitational production of vector
fields during inflation (see Refs.~\cite{VI,Lidsey}). 
This is mostly because, to achieve particle production
in a de-Sitter background, the field in question must be light enough for
its Compton wavelength to extend beyond the horizon. However, a massless vector
field is conformally invariant and, therefore, it does not couple to the 
inflating gravitational background, which means that it does not undergo 
particle production. Hence, vector field generation during inflation has been 
ignored.

In this work I investigate the possibility that a vector field with non-zero 
mass undergoes indeed particle production during inflation. My motivation was 
originally the possibility that a small, albeit non-zero, mass may lead to
something interesting. However, I have found that this is not a promising 
direction, as it is shown below. Nevertheless, I have  discovered that a 
negative mass-squared comparable to the Hubble scale can indeed result to the 
desired scale-invariant superhorizon spectrum of perturbations. 

In contrast to previous work \cite{VI,Lidsey} the vector field considered is
not assigned to the task of driving inflation. This is so in order to avoid
generating a large scale anisotropy, which is otherwise inevitable (see, 
however, Ref.~\cite{triad}). The curvature perturbations  are
produced in the same spirit as in the curvaton scenario \cite{curv}. Thus,
it is assumed that the vector field is subdominant during inflation. 
Consequently, particle production gives rise to isocurvature perturbations, 
which turn adiabatic at some point after inflation if the vector field manages 
to dominate the Universe before its decay. What I find is that an oscillating 
massive vector field does not result in a large scale anisotropy even when it 
dominates the Universe. Therefore, 
a vector field can indeed realise the curvaton scenario.

Throughout the paper I use natural units, where
\mbox{$c=\hbar=1$}. 
%and Newton's gravitational constant is
%\mbox{$8\pi G=m_P^{-2}$}, with \mbox{$m_P=2.4\times 10^{18}$GeV} being the
%reduced Planck mass. 
The signature of the metric is (1,-1,-1,-1).

\section{The equations of motion}\label{eom}

The Lagrangian density for a massive vector field with mass $m$ is
\begin{equation}
{\cal L}=-\frac{1}{4}F_{\rm \mu\nu}F^{\mu\nu}+\frac{1}{2}m^2A_\mu A^\mu\;,
\label{L}
\end{equation}
where, for an Abelian field, the field strength tensor is
\begin{equation}
F_{\mu\nu}=
%\nabla_\mu A_\nu -\nabla_\nu A_\mu =
\partial_\mu A_\nu - \partial_\nu A_\mu
%\;,
\;.
\label{F}
\end{equation}
%with $\nabla_\mu$ being the covariant derivative. 
%The Euler-Lagrange equations are:
%%
%\begin{equation}
%\nabla_\mu\left[\frac{\partial{\cal L}}{\partial(\nabla_\mu A_\nu)}\right]=
%\frac{\partial{\cal L}}{\partial A_\nu}\,.
%\label{EL}
%\end{equation}
Employing the above one obtains the field equations for the vector field:
\begin{equation}
[\partial_\mu+(\partial_\mu\sqrt{-\det[g_{\mu\nu}]})]
(\partial^\mu A^\nu-\partial^\nu A^\mu)+m^2A^\nu=0\,,
\label{FE}
\end{equation}
where \mbox{$\det[g_{\mu\nu}]$} is the determinant of the metric tensor
$g_{\mu\nu}$. 

Since we are interested in particle production during inflation we assume
that, to a good approximation, the spacetime is spatially flat homogeneous and
isotropic (we consider anisotropic expansion later). 
Hence we use the flat-FRW metric:
\begin{equation}
ds^2=dt^2-a^2(t)dx^idx^i,
\label{FRW}
\end{equation}
where $a=a(t)$ is the scale factor of the Universe, $x^i$ are Cartesian 
spatial coordinates with $i=1,2,3$ and Einstein summation is assumed. 
Employing the above metric into Eq.~(\ref{FE}) we obtain the temporal 
component \mbox{($\nu=0$)} of the field equations:
\begin{equation}
%\mbox{\boldmath $\nabla$}\cdot\mbox{\boldmath $\dot{A}$}
\mbox{\boldmath $\nabla\cdot\dot{A}$}
-\nabla^2 A_t+(am)^2 A_t=0\,,
\label{v=0}
\end{equation}
where the dot denotes derivative with respect to the cosmic time $t$ and
\mbox{\boldmath $\nabla$} stands for the divergence %or the gradient
while \mbox{$\nabla^2\equiv\partial_i\partial_i$} is the Laplacian.
Similarly, we obtain the spatial component \mbox{($\nu=i$)}:
\begin{equation}
\mbox{\boldmath $\ddot A$}+H\mbox{\boldmath $\dot A$}-a^{-2}
[\nabla^2 \mbox{\boldmath $A$} - \mbox{\boldmath $\nabla$} 
(\mbox{\boldmath $\nabla\cdot A$})]+m^2\mbox{\boldmath $A$}=
\mbox{\boldmath $\nabla$}(\dot A_t+HA_t)\,,
\label{v=i}
\end{equation}
where \mbox{$H\equiv\dot a/a$} is the Hubble parameter and in the right hand 
side of the above \mbox{\boldmath $\nabla$} denotes the gradient.

Now, contracting Eq.~(\ref{FE}) with $\partial_\nu$ 
we obtain an integrability condition, which reads
\begin{equation}
(am)^2\dot A_t-m^2\mbox{\boldmath $\nabla\cdot A$}+
3H(\nabla^2A_t-\mbox{\boldmath $\nabla\cdot\dot{A}$})=0\,.
\label{I0}
\end{equation}
Combining the above with Eq.~(\ref{v=0}) we find
\begin{equation}
\dot A_t+3HA_t-a^{-2}\mbox{\boldmath $\nabla\cdot A$}=0\,.
\label{I}
\end{equation}
Plugging this into Eq.~(\ref{v=i}) we obtain
\begin{equation}
\mbox{\boldmath $\ddot A$}+H\mbox{\boldmath $\dot A$}+m^2\mbox{\boldmath $A$}
-a^{-2}\nabla^2\mbox{\boldmath $A$}=-2H\mbox{\boldmath $\nabla$}A_t\;,
\label{EoM}
\end{equation}

We expect inflation to homogenise the vector field and, therefore,
\begin{equation}
\partial_i A_\mu=0\quad\forall\quad\mu\in[0,4]\,.
\label{hom0}
\end{equation}
Enforcing this condition into Eq.~(\ref{v=0}) we obtain 
\begin{equation}
A_t=0\,.
\label{At=0}
\end{equation}
Using Eqs.~(\ref{hom0}) and (\ref{At=0}) into Eq.~(\ref{EoM}) we find
\begin{equation}
\mbox{\boldmath $\ddot A$}+H\mbox{\boldmath $\dot A$}+m^2\mbox{\boldmath $A$}
=0\,.
\label{EoMhom}
\end{equation}
The above is reminiscent to the Klein-Gordon equation of a homogeneous scalar 
field in an expanding Universe, with the crucial difference that the friction
term does not feature a factor of 3.

We are interested in the generation of superhorizon perturbations of the
vector field, which might be responsible for the curvature perturbations in
the Universe. Therefore, we perturb the vector field around the homogeneous
value $A_\mu(t)$ as follows:
\begin{equation}
\begin{array}{l}
A_\mu(t, \mbox{\boldmath $x$})=A_\mu(t)+\delta A_\mu(t, \mbox{\boldmath $x$})
\quad\Rightarrow\\
\\
\mbox{\boldmath $A$}(t, \mbox{\boldmath $x$})=
\mbox{\boldmath $A$}(t)+\delta \mbox{\boldmath $A$}(t, \mbox{\boldmath $x$})
\;\;\&\;\;
A_t(t, \mbox{\boldmath $x$})=\delta A_t(t, \mbox{\boldmath $x$}),\!\!\!\!\!\!
\end{array}
\label{pert}
\end{equation}
where we took into account Eq.~(\ref{At=0}). In the above 
\mbox{\boldmath $A$}$(t)$ satisfies Eq.~(\ref{EoMhom}). In view of 
Eqs.~(\ref{EoMhom}) and (\ref{pert}), Eqs.~(\ref{v=0}) and (\ref{EoM}) become
\begin{eqnarray}
\hspace{-1cm} & &
\mbox{\boldmath $\nabla\cdot$}\dot{(\delta\mbox{\boldmath $A$})}
-\nabla^2 \delta A_t+(am)^2 \delta A_t=0
\label{v=0pert}\\
\hspace{-1cm} & & \nonumber\\
\hspace{-1cm} & &
\ddot{(\delta\mbox{\boldmath $A$})}+H\dot{(\delta\mbox{\boldmath $A$})}+
m^2\delta\mbox{\boldmath $A$}-a^{-2}\nabla^2\delta\mbox{\boldmath $A$}=
-2H\mbox{\boldmath $\nabla$}\delta A_t\,.
\label{EoMpert}
\end{eqnarray}

Now, let us switch to momentum space by Fourier expanding the perturbations:
\begin{equation}
\delta A_\mu(t, \mbox{\boldmath $x$})=
\int\frac{d^3k}{(2\pi)^{3/2}}\;\delta{\cal A}_\mu (t, \mbox{\boldmath $k$})\,
\exp(i\mbox{\boldmath $k\cdot x$})\,.
\label{fourier}
\end{equation}
Using the above, Eq.~(\ref{v=0pert}) becomes
\begin{equation}
\delta{\cal A}_t+
\frac{i\partial_t(\mbox{\boldmath $k\cdot$}\delta\mbox{\boldmath $\cal A$})}%
{k^2+(am)^2}=0\,,
\label{calAt}
\end{equation}
where \mbox{$k^2\equiv$}~\mbox{\boldmath $k\cdot k$}. Using this and 
Eq.~(\ref{fourier}) we can write Eq.~(\ref{EoMpert}) as
\begin{equation}
\ddot{(\delta\mbox{\boldmath $\cal A$})}+
H\dot{(\delta\mbox{\boldmath $\cal A$})}+
m^2\delta\mbox{\boldmath $\cal A$}-
\left(\frac{k}{a}\right)^2\!\delta\mbox{\boldmath $\cal A$}
+2H\frac{\mbox{\boldmath $k$}
\partial_t(\mbox{\boldmath $k\cdot$}\delta\mbox{\boldmath $\cal A$})}%
{k^2+(am)^2}=0\,.
\label{calEoM}
\end{equation}

We can rewrite the above in terms of the components parallel and perpendicular
to \mbox{\boldmath $k$}, defined as:
\begin{equation}
\delta\mbox{\boldmath $\cal A$}^\parallel\equiv
\frac{\mbox{\boldmath $k$}(\mbox{\boldmath $k\cdot$}%
\delta\mbox{\boldmath $\cal A$})}{k^2}
\quad\&\quad
\delta\mbox{\boldmath $\cal A$}^\perp\equiv
\delta\mbox{\boldmath $\cal A$}-\delta\mbox{\boldmath $\cal A$}^\parallel.
\label{perpparal}
\end{equation}
Thus, we obtain the following equations of motion for 
the vector field perturbations in momentum space:
\begin{eqnarray}
&
\left[\partial_t^2+H\partial_t+m^2+\left(\frac{k}{a}\right)^2\right]
\delta\mbox{\boldmath $\cal A$}^\perp 
=0 &
\label{EoMperp}\\
 & & \nonumber\\
& \hspace{-1cm} \left[\partial_t^2+\left(1+\frac{2k^2}{k^2+(am)^2}\right)
H\partial_t+m^2+\left(\frac{k}{a}\right)^2\right]
\delta\mbox{\boldmath $\cal A$}^\parallel=0\,. &
\label{EoMparal}
\end{eqnarray}

\section{Particle production}

To investigate particle production during inflation for the vector field
we need to solve the equation of motion for the perturbations of the field.
The integration constants are then evaluated by matching the solution to
the vacuum at early times (when \mbox{$k/aH\rightarrow+\infty$}), i.e. by
demanding
\begin{equation}
\lim_{_{\hspace{.5cm}
\frac{k}{aH}\rightarrow+\infty}}\hspace{-.5cm}
\delta{\cal A}_k=
\frac{1}{\sqrt{2k}}\exp(ik/aH),
\label{match}
\end{equation}
where \mbox{$\delta{\cal A}_k\equiv\delta$\mbox{\boldmath $\cal A$}
$(t, \mbox{\boldmath $k$})$} and we note that at early times the perturbation 
in question is well within the horizon, which means that 
\mbox{$a\rightarrow 1$} and \mbox{$k/aH\rightarrow kt$}.

Afterwards we evaluate the solution at late times, when the perturbation is 
superhorizon in size (i.e. when \mbox{$k/aH\rightarrow 0^+$}). The power 
spectrum is obtained by 
\begin{equation}
{\cal P_A}=\frac{k^3}{2\pi^2}\left|\hspace{-.5cm}\lim_{_{\hspace{.5cm}
\frac{k}{aH}\rightarrow\mbox{\scriptsize 0}^{^{+}}}}\hspace{-.5cm}
\delta{\cal A}_k\right|^2.
\label{PA}
\end{equation}
We assume that, during inflation, $H$ is constant.

\subsection{The transverse component}

Solving Eq.~(\ref{EoMperp}) and matching to the vacuum in Eq.~(\ref{match}) we
obtain the solution
\begin{equation}
\delta{\cal A}_k%^\perp
=\frac{a^{-1/2}}{1-i}\sqrt{\frac{2\pi}{H}}
\frac{e^{i\pi\nu/2}}{1-e^{i2\pi\nu}}
[J_\nu(k/aH)-e^{i\pi\nu}J_{-\nu}(k/aH)],
%\left[J_\nu\left(\frac{k}{aH}\right)
%-e^{i\pi\nu}J_{-\nu}\left(\frac{k}{aH}\right)\right],
\label{solu}
\end{equation}
where with $J_\nu$ we denote Bessel functions of the first kind and
\begin{equation}
\nu\equiv\sqrt{\frac{1}{4}-\left(\frac{m}{H}\right)^2}.
\end{equation}
The above solution at late times approaches
\begin{eqnarray}
\lim_{_{\hspace{.5cm}
\frac{k}{aH}\rightarrow\mbox{\scriptsize 0}^{^{+}}}}\hspace{-.5cm}
\delta{\cal A}_k & = & \frac{a^{-1/2}}{1-i}\sqrt{\frac{2\pi}{H}}
\frac{e^{i\pi\nu/2}}{1-e^{i2\pi\nu}}\;\times\nonumber\\
 & & \nonumber\\
 & & \hspace{-2.5cm}\times
\left[\frac{1}{\Gamma(1+\nu)}\left(\frac{k}{2aH}\right)^\nu-
\frac{e^{i\pi\nu}}{\Gamma(1-\nu)}\left(\frac{k}{2aH}\right)^{-\nu}
\right].
\label{late1}
\end{eqnarray}
Hence, using Eq.~(\ref{PA}) we find that the dominant contribution to the 
power spectrum is
\begin{equation}
{\cal P_A}\approx\frac{8\pi|\Gamma(1-\nu)|^{-2}}{(1-\cos 2\pi\nu)}
\left(\frac{aH}{2\pi}\right)^2\left(\frac{k}{2aH}\right)^{3-2\nu}
\label{PAperp}
\end{equation}

Now, considering a light vector field with \mbox{$m\ll H$} we see that
\mbox{$\nu\approx\frac{1}{2}-(m/H)^2$}. As a result the dominant term in the 
power spectrum is simply the vacuum value 
\begin{equation}
{\cal P_A}^{\rm vac}=\left(\frac{k}{2\pi}\right)^2,
\label{PAvac}
\end{equation}
which agrees with the expectations, since, when \mbox{$m\rightarrow 0$}
the vector field becomes conformally invariant and, therefore, it is not 
gravitationally produced because it does not couple to the expanding 
gravitational background. In fact, for \mbox{$m\ll H$}, the largest 
contribution to the power spectrum due to a non-zero mass is
\begin{equation}
\delta{\cal P_A}=2\pi\left(\nu-\frac{1}{2}\right)\left(\frac{aH}{2\pi}\right)^2
\left(\frac{k}{2aH}\right)^3,
\label{dP}
\end{equation}
which is subdominant to the vacuum value for superhorizon scales.

However, if the field is not effectively massless but instead we
have
\begin{equation}
m^2\approx-2H^2\quad\Rightarrow\quad\nu\approx 3/2\,,
\label{m}
\end{equation}
then we find that a scale invariant spectrum of perturbations is indeed 
recovered with
\begin{equation}
{\cal P_A}\approx a^2\left(\frac{H}{2\pi}\right)^2
\label{PAflat}
\end{equation}
as in the case of a massless scalar field. We will discuss the
reason for this result in Sec.~\ref{phys}. For the time being we note that
the transverse component of a light vector field cannot be gravitationally 
generated during inflation.

\subsection{The longitudinal component}

Turning our attention now to $\delta{\cal A}_k^\parallel$, the first thing to
point out is that Eq.~(\ref{EoMparal}) is impossible to solve analytically.
One can only approximate the solutions in some extreme cases.

Consider first that \mbox{$k\ll am$}. Then it is evident that the equation 
assumes the same form as the case of $\delta{\cal A}_k^\perp$ and, therefore,
the results are identical to the previous section. However, in the opposite 
case, when \mbox{$k\gg am$}, the equation becomes
\begin{equation}
\left[\partial_t^2+3H\partial_t+m^2+\left(\frac{k}{a}\right)^2\right]
\delta\mbox{\boldmath $\cal A$}^\parallel\simeq 0\,,
\label{long}
\end{equation}
which, in fact, is identical to the equation of motion for a perturbation
of a light scalar field in a de-Sitter background. The solution of the above, 
after matching to the vacuum in Eq.~(\ref{match}),
%we obtain 
%%
%\begin{equation}
%\delta{\cal A}_k^\parallel=\frac{a^{-1/2}}{1-i}\sqrt{\frac{2\pi}{H}}
%\frac{e^{i\pi\nu/2}}{1-e^{i2\pi\nu}}[J_\nu(k/aH)-e^{i\pi\nu}J_{-\nu}(k/aH)],
%\label{solu1}
%\end{equation}
%which 
is of identical form to Eq.~(\ref{solu}) with the crucial difference
that, this time
\begin{equation}
\nu\equiv\sqrt{\frac{9}{4}-\left(\frac{m}{H}\right)^2}.
\label{nu}
\end{equation}
Thus, the dominant contribution to the power spectrum is again given by 
Eq.~(\ref{PAperp}). 

Considering, therefore, an effectively massless field with \mbox{$m\ll H$} we
have 
\begin{equation}
\nu\approx\frac{3}{2}-\eta\quad{\rm with}\quad
\eta\equiv\frac{1}{3}\left(\frac{m}{H}\right)^2.
\end{equation}
In view of the above the power spectrum in Eq.~(\ref{PAperp}) becomes
\begin{equation}
{\cal P_A}\approx a^2\left(\frac{H}{2\pi}\right)^2
\left(\frac{k}{2a}\right)^{2\eta},
\label{PAlong}
\end{equation}
which is approximately scale invariant approaching the value of 
Eq.~(\ref{PAflat}) for extremely light fields. Parameterising the scale 
dependence of the perturbations in the usual manner
\begin{equation}
{\cal P_A}(k)\propto k^{n_s-1},
\label{PAns}
\end{equation}
we obtain for the spectral index the result
\begin{equation}
n_s=1+2\eta\,,
\label{ns}
\end{equation}
which is the usual finding in the case of a light scalar field.\footnote{There 
is no contribution from \mbox{$\epsilon\equiv-\dot H/H^2$} to the spectral 
index because we have taken \mbox{$H=$ const}.}

The similarity
between the longitudinal component of a massive vector field to a scalar field
can be understood if one considers that the mass of the vector field is
due to the Higgs mechanism, in which case the longitudinal component 
corresponds to the scalar degree of freedom (the Goldstone boson) which has 
been ``consumed'' by the vector field in the Higgs process. However, we should 
also note here that, in the limit of \mbox{$m\rightarrow 0$} the longitudinal 
component of the vector field becomes unphysical.
(The superhorizon spectra of $\delta{\cal A}_k^\perp$ and 
$\delta{\cal A}_k^\parallel$ in the case when \mbox{$m\ll H$} are shown in
Figure~1.)

Thus, we see that a light vector field may indeed obtain an almost scale 
invariant superhorizon perturbation spectrum, through its longitudinal 
component. However, the price to pay is that the condition \mbox{$k\gg am$}
has to be satisfied throughout the superhorizon evolution of the perturbations
in question. This means that, for the cosmological scales, we must satisfy the
constraint
\begin{equation}
a_*H_*\equiv k_*>ma_{\rm end}\Rightarrow m<e^{-N_*}H_*\;,
\label{mcons}
\end{equation}
where `*' denotes the epoch when the cosmological scales exit the horizon
during inflation and `end' denotes the end of inflation, while $N$ corresponds
to the number of remaining e-foldings of inflation. For the cosmological scales
we have \mbox{$N_*\gsim 45$} and also, typically \mbox{$H_*\lsim 10^{13}$GeV}.
Hence, we find that the mass of the vector field has to be 
\mbox{$m\lsim{\cal O}(100)$ eV}. Such an extremely light field cannot decay
before nucleosynthesis and, therefor, cannot be responsible for the curvature 
perturbations imprinted on the thermal bath of the Hot Big Bang. The only 
possible solution would be to consider a subsequent period of inflation, 
occurring after the decay of the vector field, which would last long enough to
drastically diminish $N_*$, but not too long to render the original 
inflationary period irrelevant. This highly contrived scenario, however, 
suffers from another problem, as will be shown in the next section.

\begin{figure}[t]
\includegraphics[width=85mm,angle=0]{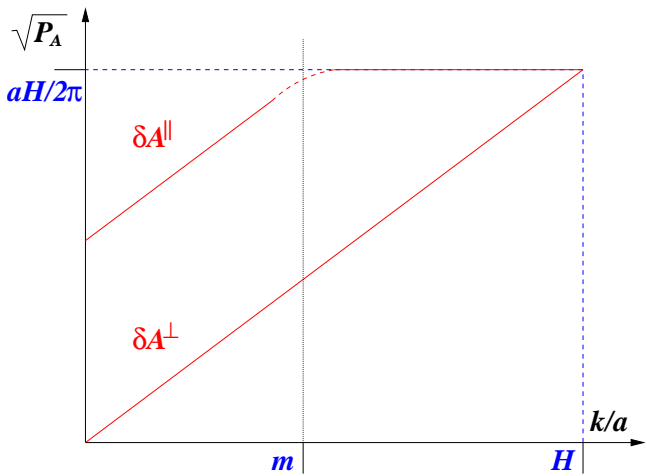}
\caption{
Illustration of the superhorizon spectra of the transverse and longitudinal
components of the perturbation of a light vector field. The transverse 
component corresponds to the vacuum spectrum. The longitudinal component mimics
the transverse component at small momenta but at large momenta its spectrum 
becomes approximately scale invariant.}
\end{figure}

\section{Evolution during and after Inflation}

As we mentioned in the introduction, in order to avoid large scale anisotropy 
in the Universe, we do not consider the possibility that the vector field in
question can act as the inflaton field. Instead we assume that the vector field
may imprint its spectrum of curvature perturbations onto the Universe at some
time after the end of inflation, when it becomes dominant (or nearly dominant)
before its decay. In other words, we assume that our vector field acts as a
curvaton field. In this case the amplitude of the curvature perturbations is
determined by the dynamics of the field after inflation as well. 

\subsection{The evolution of the vector field}

Assume that the homogenised vector field lies along the $z$-direction
$$
A_\mu=(0,0,0,A_z(t)\,)\,.
$$
Now, $A_z(t)$ satisfies Eq.~(\ref{EoMhom}). Solving this equation during
inflation, where \mbox{$H\simeq$ const.} we obtain
\begin{eqnarray}
A_z(t) & = & \left[-\frac{\dot A_z(0)}{HR}
-\frac{1-R}{2R}A_z(0)\right]
e^{-\frac{1}{2}H\Delta t(1+R)}+\nonumber\\
 & + & 
\left[\frac{\dot A_z(0)}{HR}
+\frac{1+R}{2R}A_z(0)\right]
e^{-\frac{1}{2}H\Delta t(1-R)},
\label{A1}
\end{eqnarray}
%
%\begin{eqnarray}
%A_z(t) & = & C_1\exp\left\{-\frac{1}{2}H\Delta t
%\left[1+\sqrt{1-4\left(\frac{m}{H}\right)^2}\right]\right\}+\nonumber\\
% & + & C_2\exp\left\{-\frac{1}{2}H\Delta t
%\left[1-\sqrt{1-4\left(\frac{m}{H}\right)^2}\right]\right\},
%\label{A1}
%\end{eqnarray}
%where
%\begin{eqnarray}
%\hspace{-.7cm}
%C_1 & = & -\frac{\dot A_z(0)}{H\sqrt{1-4\left(\frac{m}{H}\right)^2}}
%-\frac{1-\sqrt{1-
%4\left(\frac{m}{H}\right)^2}}{2\sqrt{1-4\left(\frac{m}{H}\right)^2}}A_z(0)
%\label{C1}\\
%\hspace{-.7cm}
%C_2 & = & \frac{\dot A_z(0)}{H\sqrt{1-4\left(\frac{m}{H}\right)^2}}
%+\frac{1+\sqrt{1-
%4\left(\frac{m}{H}\right)^2}}{2\sqrt{1-4\left(\frac{m}{H}\right)^2}}A_z(0)\,,
%\label{C2}
%\end{eqnarray}
where \mbox{$R\equiv\sqrt{1-4\left(\frac{m}{H}\right)^2}$},
$\Delta t$ is the elapsed time and  `(0)' denotes the initial value.
When \mbox{$m\ll H$} the above give
\begin{eqnarray}
\!\!\!\!A_z(t) & \simeq & 
\left[\frac{\dot A_z(0)}{H}+A_z(0)\right]
\exp\left[-H\Delta t(\frac{m}{H})^2\right]-\nonumber\\
 & - & \left[\frac{\dot A_z(0)}{H}+\left(\frac{m}{H}\right)^2A_z(0)\right]
\exp\left(-H\Delta t\right)\,.
\label{Ainf}
\end{eqnarray}
Therefore, for a light vector field during inflation we find
\mbox{$A_z\rightarrow A_z(0)+\dot A_z(0)/H=$ const.}

After the end of inflation solving Eq.~(\ref{EoMhom}) we obtain
\begin{equation}
A_z(t)=t^x[c_1J_x(mt)+c_2Y_x(mt)]\,,
\label{AD}
\end{equation}
%
%\begin{equation}
%A_z(t)=t^{1/6}[c_1J_{1/6}(mt)+c_2Y_{1/6}(mt)]\,,
%\label{AMD}
%\end{equation}
%for matter domination and
%%
%\begin{equation}
%A_z(t)=t^{1/4}[c_1J_{1/4}(mt)+c_2Y_{1/4}(mt)]\,,
%\label{ARD}
%\end{equation}
%for radiation domination, 
where $c_1,c_2$ are constants of integration,
$Y_x$ is a Bessel function of the second kind and 
\begin{equation}
x\equiv\frac{1+3w}{6(1+w)}
\label{x}
\end{equation}
with $w$ being the barotropic parameter (\mbox{$w=0$} 
\{\mbox{$w=\frac{1}{3}$}\} for matter \{radiation\} domination).

When \mbox{$m\ll H$} we find
\begin{equation}
A_z(t)\simeq t^x\left[\frac{c_1}{\Gamma(x+1)}
\left(\frac{mt}{2}\right)^x\!\!\!\!\!\!-
c_2\frac{\Gamma(x)}{\pi}\left(\frac{mt}{2}\right)^{-x}\right].
\label{A1D}
\end{equation}
%%
%\begin{equation}
%\,\!\hspace{-.1cm}
%A_z(t)\simeq t^{1/6}\left[\frac{c_1}{\Gamma(7/6)}
%\left(\frac{mt}{2}\right)^{1/6}\!\!\!\!\!\!-
%c_2\frac{\Gamma(1/6)}{\pi}\left(\frac{mt}{2}\right)^{-1/6}\right]
%\hspace{-.5cm}
%\label{A1MD}
%\end{equation}
%for matter domination and
%%
%\begin{equation}
%\,\!\hspace{-.1cm}
%A_z(t)=t^{1/4}\left[\frac{c_1}{\Gamma(5/4)}
%\left(\frac{mt}{2}\right)^{1/4}\!\!\!\!\!\!-
%c_2\frac{\Gamma(1/4)}{\pi}\left(\frac{mt}{2}\right)^{-1/4}\right]
%\hspace{-.5cm}
%\label{A1RD}
%\end{equation}
%for radiation domination. 
%
From the above, since \mbox{$(mt/2)\sim m/H\ll 1$}, 
we see that, in both cases, \mbox{$A_z(t)\simeq$ const.} Thus, we have 
verified that both during and after inflation, when \mbox{$m\ll H$} the vector
field is overdamped and remains frozen.

Now, consider the opposite case \mbox{$m\gg H$}. In this case, from
%Eqs.~(\ref{AMD}) and (\ref{ARD}) 
Eq.~(\ref{AD})
we have
\begin{eqnarray}
%\hspace{-.5cm}
A_z(t) & \simeq & t^{-\frac{1}{3(1+w)}}\sqrt{\frac{2}{\pi m}}%\;\times
\left\{c_1\cos\left[mt-\frac{\pi}{2}\left(x+\frac{1}{2}\right)\right]\right.+
\nonumber\\
 & & \hspace{1.6cm}+
\left.c_2\sin\left[mt-\frac{\pi}{2}\left(x+\frac{1}{2}\right)\right]\right\}\!.
\label{A2D}
\end{eqnarray}
%%
%\begin{eqnarray}
%\hspace{-.5cm}
%A_z(t) & \simeq & t^{-1/3}\sqrt{\frac{2}{\pi m}}\;\times
%\nonumber\\
% & \times & 
%\left[c_1\cos\left(mt-\frac{\pi}{3}\right)+
%c_2\sin\left(mt-\frac{\pi}{3}\right)\right]
%\label{A2MD}
%\end{eqnarray}
%for matter domination and
%%
%\begin{eqnarray}
%\hspace{-.5cm}
%A_z(t) & \simeq & t^{-1/4}\sqrt{\frac{2}{\pi m}}\;\times
%\nonumber\\
% & \times & 
%\left[c_1\cos\left(mt-\frac{3\pi}{8}\right)+
%c_2\sin\left(mt-\frac{3\pi}{8}\right)\right]
%\label{A2RD}
%\end{eqnarray}
%for radiation domination. 
%
Hence, we see that, when the vector field becomes
heavy then it engages in damped harmonic oscillations with envelope decreasing 
as
\begin{equation}
\overline{A_z(t)}\propto a^{-1/2}.
\label{envelope}
\end{equation}

\subsection{The energy momentum tensor}

Now that we know how the vector field evolves during and after inflation 
we can compute if and when it will come to dominate the Universe, in order
to imprint its superhorizon perturbation spectrum. To find this
we follow the evolution of the energy-momentum tensor of the vector field.

Using Eq.~(\ref{L}), the energy momentum tensor for $A_\mu$ is
\begin{eqnarray}
& T_{\mu\nu}=
\frac{1}{4}g_{\mu\nu}F_{\rho\sigma}F^{\rho\sigma}-
F_{\mu\rho}F_\nu^{\;\rho}+ & \nonumber\\
& \hspace{1.5cm}
+\,m^2\left(A_\mu A_\nu-\frac{1}{2}g_{\mu\nu}A_\rho A^\rho\right). & 
\label{Tmn}
\end{eqnarray}
In order to imprint its perturbation spectrum onto the Universe the 
vector field must dominate (or nearly dominate) the Universe, in accordance to
the curvaton scenario. Since this may result into large scale anisotropy we 
consider the following metric
\begin{equation}
ds^2=dt^2-a^2(t)(dx^2+dy^2)-b^2(t)dz^2,
\label{metric}
\end{equation}
where $b(t)$ is the scale factor along the $z$-direction. 
In view of the above,
the energy-momentum tensor can be written in the form
\begin{equation}
T_\mu^{\,\nu}={\rm diag}(\rho_A, -p_\perp, -p_\perp, +p_\perp)\,,
\label{Tdiag}
\end{equation}
where
\begin{equation}
\rho_A\equiv\rho_{\rm kin}+V\qquad
p_\perp\equiv\rho_{\rm kin}-V
\label{rp}
\end{equation}
with
\begin{eqnarray}
\rho_{\rm kin} & \equiv & -\frac{1}{4}F_{\mu\nu}F^{\mu\nu}
\;=\;\frac{1}{2b^2}\dot A_z^2\label{rkin}\\
 & & \nonumber\\
V & \equiv & -\frac{1}{2}m^2A_\mu A^\mu
\;=\;\frac{1}{2b^2}m^2A_z^2.\label{V}
\end{eqnarray}

From Eq.~(\ref{Tdiag}) we see that the energy momentum tensor for our vector
field resembles the one of a perfect fluid, with the crucial difference that
the pressure along the longitudinal direction is of opposite sign to the 
pressure along the transverse directions. Thus, if the pressure is non-zero
and the vector field dominates the Universe, then large scale anisotropy will 
be generated. This is the reason we did not consider that $A_\mu$ can play the
role of the inflaton field in the first place.\footnote{%
In the limit \mbox{$m\rightarrow 0$} it is easy to see that $T_{\mu\nu}$
becomes traceless as it should. Also, \mbox{$\rho_A\rightarrow\rho_{\rm kin}$}.
In the case of isotropic expansion ($b=a$) (i.e. when $A_\mu$ is subdominant)
the solution of Eq.~(\ref{EoMhom}) suggests, \mbox{$\dot A_z\propto a^{-1}$}. 
Hence, \mbox{$\rho_A=\frac{1}{2}(\dot A_z/a)^2\propto a^{-4}$}, which is the 
conformal invariant result for radiation.}

To study the evolution of the energy density of the vector field
we begin by assuming that originally $\rho_A$ is subdominant, in an isotropic
Universe. In this case we can employ Eq.~(\ref{EoMhom}), which can be written 
as
\begin{equation}
\ddot A_z+m^2A_z\left[1+\left(\frac{H}{m}\right)^2\frac{\dot A_z}{HA_z}\right]
=0\,.
\end{equation}
When \mbox{$m\gg H$} we have shown that $A_z$ oscillates harmonically
with envelope decreasing as shown in Eq.~(\ref{envelope}). Using this we can
approximate \mbox{$\dot A_z/HA_z\approx\dot{\bar A}_z/H\bar A_z=-\frac{1}{2}$}.
Inserting this into the above and considering that \mbox{$m\gg H$} we obtain
\begin{equation}
\ddot A_z+m^2A_z\approx 0\,,
\label{harm}
\end{equation}
which verifies that, since the oscillation period is much smaller that
the Hubble time, the oscillations are practically harmonic. This means that,
on average,
\begin{equation}
\overline{\dot A_z^2}=m^2\overline{A_z^2}
\label{average}
\end{equation}
Now, Eq.~(\ref{EoMhom}) can also be written as
\begin{equation}
\frac{d}{dt}\left(\frac{1}{2}\dot A_z^2+\frac{1}{2}m^2A_z^2\right)
+H\dot A_z^2=0\,.
\end{equation}
In view of Eq.~(\ref{average}) the above is recast as
\begin{equation}
\frac{d}{dt}(\overline{\dot A_z^2})+H\overline{\dot A_z^2}=0\,.
\label{aveq}
\end{equation}
From Eqs.~(\ref{rp}), (\ref{rkin}) and (\ref{V}), considering also 
Eq.~(\ref{average}) we have
\begin{equation}
\overline{\dot A_z^2}=a^2\overline{\rho_A}\,.
\end{equation}
Hence, Eq,~(\ref{aveq}) suggests that
\begin{equation}
\overline{\rho_A}\propto a^{-3}.
\label{rA}
\end{equation}
This means that, when the vector field begins oscillating, its density
scales as pressureless matter. This is not surprising because 
Eq.~(\ref{average}) implies that, on average 
\mbox{$\overline{\rho_{\rm kin}}=\overline V$}, which means that 
\mbox{$\overline{p_\perp}=0$}. Hence, all the pressure components in 
Eq.~(\ref{Tdiag}) are equal to zero. The result in Eq.~(\ref{rA}) is also 
obtained by considering that 
\begin{equation}
\overline{\rho_A}=2\overline V=a^{-2}m^2\overline{A_z^2}\propto a^{-3},
\label{rbar}
\end{equation}
where we also used Eq.~(\ref{envelope}). 

The above behaviour does not change if the Universe becomes anisotropic.
If this is so, then it is easy to verify that, using the metric in 
Eq.~(\ref{metric}) and Eq.~(\ref{FE}), the equivalent to Eq.~(\ref{EoMhom}) is
\begin{equation}
\ddot A_z+(2H_a-H_b)\dot A_z+m^2A_z=0\,,
\end{equation}
where \mbox{$H_a\equiv\dot a/a$} and \mbox{$H_b\equiv\dot b/b$}. Considering
\mbox{$m\gg H_a,H_b$} one arrives at Eq.~(\ref{average}), from which all the 
results follow.

The fact that the average pressure of an oscillating vector field is zero
along both the longitudinal and the transverse directions implies that
{\em when the density of the vector field dominates the Universe this does
not cause anisotropic expansion}. Hence, we can use the curvaton mechanism
to imprint the superhorizon perturbation spectrum onto the Universe without
the danger of generating a large scale anisotropy, provided that the
vector field is oscillating at domination.

\subsection{Generating the curvature perturbation}

The curvaton mechanism transforms the spectrum of perturbations of the vector 
field into a spectrum of curvature perturbations as follows. As discussed by
Lyth and Wands in Ref.~\cite{curv}, on a foliage of spacetime corresponding to 
spatially flat hypersurfaces, the curvature perturbation
attributed to each of the Universe components (labelled by the index $n$) is 
given by
\begin{equation}
\zeta_n\equiv-H\frac{\delta\rho_n}
{\dot{\rho}_n},
\label{zetai}
\end{equation}
where $\rho_n$ and $\delta\rho_n$ are, respectively, the density and its 
perturbation of the component in question.

The total curvature perturbation $\zeta(t)$, which is also given by 
Eq.~(\ref{zetai}) with $\rho_n$ and $\delta\rho_n$ replaced, respectively, by 
the total density of the Universe $\rho=\sum_n\rho_n$ and its perturbation
$\delta\rho$, may be calculated as follows. Using the fact that 
\mbox{$\delta\rho=\sum_n\delta\rho_n$} and the continuity equation
\mbox{$\dot{\rho}_n=-3H(\rho_n+p_n)$}, where $p_n$ is the pressure of the 
$n$-th component of the Universe, it is easy to find that
\begin{equation}
\zeta=\sum_n\frac{\rho_n+p_n}{\rho+p}\;\zeta_n,
\label{zeta0}
\end{equation}
where $p=\sum_np_n$ is the total pressure. Now, since in the curvaton scenario,
all contributions to the curvature perturbation other than the curvaton's are 
negligible, we find that
\begin{equation}
\zeta=\zeta_A
\left(\frac{1}{1+w}\right)_{\rm dec}
\left.\frac{\rho_A}{\rho}
\right|_{\,\rm dec},
\label{z}
\end{equation}
where \mbox{$\zeta\simeq 2\times 10^{-5}$} is the observed curvature 
perturbation and $\zeta_A$ is the partial curvature perturbation 
of the vector field, for which we have assumed that the pressure is zero,
when it is oscillating as previously discussed. The right hand side of this 
equation is evaluated at the time when the vector field decays and this is 
indicated by the subscript `dec'. 
If the vector field decays after it dominates the Universe then 
\mbox{$\rho\rightarrow \rho_A$}, which means that \mbox{$\zeta\approx\zeta_A$}.

To obtain $\zeta_A$ we employ Eq.~(\ref{zetai}) and the continuity equation, 
which suggest
\begin{equation}
\zeta_A=\left.\frac{\delta\rho_A}{3\rho_A}\right|_{\rm dec}\simeq
\frac{2}{3}\left.\frac{\delta\bar{A_z}}{\bar{A_z}}\right|_{\rm dec}
\approx\frac{2}{3}\left.\frac{\delta A_z}{A_z}\right|_*\;,
\label{zA}
\end{equation}
where the bar denotes the amplitude of the oscillating vector field at the 
time of decay [c.f. Eq.~(\ref{rbar})]. 
In the above we have considered that, after the end of 
inflation, the ratio $\delta A_z/A_z$ remains constant because either the
vector field and its perturbation are frozen (when $m\ll H$) or they are
oscillating (when $m\gg H$), with the same equation of motion. This is indeed 
so since the perturbations in question are superhorizon in size when the vector
field decays, so spatial gradients in Eq.~(\ref{EoMpert}) are negligible and, 
hence, Eq.~(\ref{EoMpert}) becomes of identical form to Eq.~(\ref{EoMhom}).

Now, in Eq.~(\ref{zA}) $A_z(t)$ is the homogeneous zero-mode, which has no 
$k$-dependence. This means that the $\zeta_A(k)$ has the same $k$-dependence as
the perturbation $\delta A_z(k)$. Hence, {\em a scale-invariant superhorizon 
spectrum of vector field perturbations can give rise to a scale-invariant 
superhorizon curvature perturbation spectrum} provided the vector field 
is heavy and engages into oscillations before it dominates the Universe.

The above are in stark contrast to the case when the vector field dominates
while it is still light. As we have shown, in this case $A_z$ is frozen to
a constant value, which suggests that \mbox{$\rho_{\rm kin}\approx 0$} and
\mbox{$p_\perp=-\rho_A=-V$}. This implies that the Universe inflates
along the transverse direction but not along the longitudinal direction
because the longitudinal pressure is positive (c.f. Eq.~(\ref{Tdiag})).
Hence, when \mbox{$m\ll H$} we have to demand that $\rho_A$ is subdominant
in order to retain isotropy.

We should note here that, for a light vector field, $\rho_A$ is {\em not} 
constant. Indeed, while \mbox{$A_z\simeq$ const.}, we see from Eq.~(\ref{V})
that \mbox{$\rho_A=V\propto a^{-2}$} (with $a=b$). Hence, despite the fact
that the vector field is frozen, its density decreases as $a^{-2}$. This 
undermines even further the possibility of using a light vector field to
generate the curvature perturbation, because its density during inflation
is exponentially suppressed. Bearing in mind that the field needs to dominate
(or nearly dominate) after inflation and decay before nucleosynthesis while
having a tiny mass due to the bound in Eq.~(\ref{mcons}) it is easy to see that
a successful scenario is rather unlikely (in fact it is inviable). However, if 
we abandon the light field assumption we might be able to 
attain the desired result as we discuss in the next section.

\section{The ``physical'' vector field}\label{phys}

The careful reader might have been alarmed by the fact that the results in
Eqs.~(\ref{PAflat}) or (\ref{PAlong}) appear to be proportional to powers
of the scale factor. Similarly, the components of the energy momentum tensor
as shown in Eqs.~(\ref{rkin}) and (\ref{V}) also appear to bear an explicit
dependence to the scale factor $b$. Both these findings are not expected
because quantities such as the power spectrum or the density and
pressure are observables and should not depend on the normalisation of the 
scale factor. The problem is overcome if one realises that $A_\mu$ is more
like a `comoving' quantity, which has the Universe expansion factored out.
As implied by the form of $\cal P_A$ in Eqs.~(\ref{PAflat}) and (\ref{PAlong})
and also on the form of $V$ in Eq.~(\ref{V}) one can assume that, in an 
isotropic Universe, the spatial components of the ``physical'' vector field 
may be defined as%
\footnote{Eq.~(\ref{U}) is also motivated as follows. Suppose that $m=0$. 
Then the non-zero components of the
field strength tensor can be expressed in terms of an electric and a 
magnetic field $E_i$ and $B_i$ respectively as:
%$$
%F_{\mu\nu}=\left(
%\begin{array}{cccc}
%0      & E_x    & E_y    & E_z\\
%-E_x   & 0      & -B_z   & B_y\\
%-E_y   & B_z    & 0      & -B_x\\
%-E_z   & -B_y   & B_x    & 0\\
%\end{array},
%\right)
%$$
%which means that 
\mbox{$F_{0i}=E_i$} and \mbox{$F_{ij}=-\epsilon_{ijk}B_k$},
where $\epsilon_{ijk}$ is the totally anti-symmetric Levi-Civita tensor.
In this case the energy momentum tensor in Eq.~(\ref{Tmn}) suggests that
the energy density of the field is:
$$\rho_A\equiv T^{00}=\frac{1}{2}
\left(\frac{E_iE_i}{a^2}+\frac{B_iB_i}{a^4}\right).$$
Since, the energy density is a physical quantity we realise that the physical
electric field is $E_i/a$ and the physical magnetic field is $B_i/a^2$.
Now, $B_i$ is defined as \mbox{$B_i\equiv\epsilon_{ijk}\partial_jA_k$}. This 
means
$$
\mbox{\boldmath $B$}\equiv\mbox{\boldmath $\nabla$}\times\mbox{\boldmath $A$}
\quad\Rightarrow\quad
\mbox{\boldmath $B$}/a^2=
\mbox{\boldmath $\hat\nabla$}\times\mbox{\boldmath $V$},
$$
where the hat denotes derivatives with respect to physical coordinates 
\mbox{$r^i\equiv ax^i$} in contrast to comoving coordinates $x^i$ 
[c.f. Eq.~(\ref{FRW})], 
such that \mbox{$\hat\partial_i=\partial_i/a$}. Hence we see that, since 
\mbox{\boldmath $B$}/$a^2$ is a physical quantity, so is \mbox{\boldmath $V$}.
%
%Similarly, one can also deduce that 
%%
%$$
%\mbox{\boldmath $E$}/a=H\mbox{\boldmath $V$}+\mbox{\boldmath $\dot V$}-
%\mbox{\boldmath $\hat\nabla$}V_t\;,
%$$
%which confirms that \mbox{$V_t=A_t$}.
}
\begin{equation}
V_i\equiv A_i/a(t)\,.
%\,\quad{\rm and}\quad V_t\equiv A_t.
\label{U}
\end{equation}

%For a homogeneous $A_\mu$ we have found that \mbox{$A_t=0$} (c.f. 
%Eq.~(\ref{At=0})), which means also that a homogeneous $V_\mu$
%corresponds to \mbox{$V_t=0$}. 
%
From Eqs.~(\ref{EoMhom}) and (\ref{U}) we find that 
the spatial components 
$V_i(t)$ during inflation satisfy the equation
\begin{equation}
\mbox{\boldmath $\ddot V$}+3H\mbox{\boldmath $\dot V$}+
(2H^2+m^2)\mbox{\boldmath $V$}=0\,,
\label{EoMU}
\end{equation}
where we considered \mbox{$H\simeq$ const.} After perturbing $V_\mu$ around
the homogeneous value and following the same procedure as in Sec.~\ref{eom}
we obtain the equations of motion for the transverse component of the 
perturbations of the vector field:
\begin{equation}
\left[\partial_t^2+3H\partial_t+2H^2+m^2+\left(\frac{k}{a}\right)^2\right]
\delta\mbox{\boldmath $\cal V$}^\perp= 0\,,
\label{Uperp}
\end{equation}
where 
\begin{equation}
\delta V_\mu(t, \mbox{\boldmath $x$})=
\int\frac{d^3k}{(2\pi)^{3/2}}\;\delta{\cal V}_\mu (t, \mbox{\boldmath $k$})\,
\exp(i\mbox{\boldmath $k\cdot x$})\,,
\label{fourier1}
\end{equation}
i.e. \mbox{$\delta{\cal V}_k=\delta{\cal A}_k/a(t)$}. From 
Eqs.~(\ref{EoMU}) and (\ref{Uperp}) it is evident that the equation of motion
for the transverse
component of the ``physical'' vector field is remarkably similar to a scalar 
field (notice the factor of 3H in the friction term of Eq.~(\ref{EoMU})) of 
mass-squared
\begin{equation}
\tilde m^2=2H^2+m^2.
\label{mtild}
\end{equation}
Hence, we expect that particle production will take place when 
\mbox{$\tilde m\ll H$}, i.e. in the case when
\begin{equation}
m^2\approx -2H^2
\label{mH}
\end{equation}

Indeed, we have already seen that, when the condition in Eq.~(\ref{mH}) is
valid, the transverse component of the vector field obtains a scale 
invariant superhorizon spectrum of perturbations. In view of Eq.~(\ref{PAflat})
we see that
\begin{equation}
{\cal P_V}\approx \left(\frac{H}{2\pi}\right)^2
\label{PAflatU}
\end{equation}
exactly as in the case of a scalar field, where we used that 
\mbox{${\cal P_V}=a^{-2}{\cal P_A}$} as is evident by Eq.~(\ref{PA}) and the 
fact that \mbox{$\delta{\cal V}_k=\delta{\cal A}_k/a$}. 

Now, if we employ the condition in Eq.~(\ref{mH}) into Eq.~(\ref{A1}) 
%(\ref{C1}) and (\ref{C2}) 
we find that, during inflation, \mbox{$R=3$} and so
\begin{eqnarray}
A_z(t) & \simeq & \frac{1}{3H}\left\{
\left[2HA_z(0)+\dot A_z(0)\right]e^{H\Delta t}\right.\;+\nonumber\\
 & & \hspace{.5cm}+\;\left.\left[HA_z(0)-\dot A_z(0)\right]e^{-2H\Delta t}
\right\}.
\end{eqnarray}
Hence, after a Hubble time we see that 
\mbox{$A_z\propto e^{H\Delta t}\propto a$}, which means that 
\mbox{$V_z\equiv A_z/a\simeq$ const.} [c.f. Eq.~(\ref{U})] and, therefore the 
``physical'' vector field remains frozen during inflation. 

Expressing the components of $T_{\mu\nu}$ in terms of $V_z$ we obtain
\begin{eqnarray}
\rho_A & = & \frac{1}{2}[\dot V_z^2+2HV_z\dot V_z+(H^2+m^2)V_z^2]
\label{rU}\\
p_\perp & = & -\frac{1}{2}[\dot V_z^2+2HV_z\dot V_z+(H^2-m^2)V_z^2].
\label{pU}
\end{eqnarray}
During inflation, when we require Eq.~(\ref{mH}) to hold, we find
\begin{equation}
\rho_A=-\frac{1}{2}H^2V_z^2\quad{\rm and}\quad
p_\perp=-\frac{3}{2}H^2V_z^2,
\end{equation}
where we have used that \mbox{$V_z\simeq$ const.} during inflation.
Since $V_z$ is frozen during inflation, both the energy density and the 
pressure are constant. The fact that the energy density appears negative is
not surprising; according to Eq.~(\ref{mH}) the mass of the vector field is
tachyonic, which is a similar situation to the case when a scalar field is
placed on top of a potential hill.

\section{Vector Curvaton in Supergravity}

From the above we see that we may have a chance to generate a scale invariant
superhorizon spectrum of perturbations through the use of a vector field, 
provided the condition in Eq.~(\ref{mH}) holds during inflation, at least when
the cosmological scales exit the horizon. How can we achieve this?

In particle physics, vector fields obtain masses through the Higgs mechanism.
The masses are due to the kinetic term of the Higgs field $\phi$
\begin{equation}
{\cal L}_{D\phi}=D_\mu\phi(D^\mu\phi)^*\,,
\label{LD}
\end{equation}
where `*' here denotes charge conjugation and
\begin{equation}
D_\mu\equiv\partial_\mu+igA_\mu
\end{equation}
is the covariant derivative with $g$ being the gauge coupling. 
This means that the Higgs kinetic term in 
Eq.~(\ref{LD}) generates a mass term for the vector field $A_\mu$:
\begin{equation}
{\cal L}_m=g^2|\phi|^2A_\mu A^\mu.
\label{Lm}
\end{equation}

From this term it is in principle possible to obtain a mass of order $H$ during
inflation for the vector field. Indeed, if the Higgs field is light during 
inflation then particle production would generate a condensate of magnitude
\begin{equation}
\langle\phi^2\rangle\sim H^2\min\left\{H\Delta t, H^2/m_\phi^2\right\},
\label{phi2}
\end{equation}
where $m_\phi\ll H$ is the mass of the Higgs field and 
\mbox{$\Delta N=H\Delta t$} is the number of elapsed e-foldings of inflation.
The second term in the brackets above corresponds to the Bunch-Davis result
\cite{BD}, whereas the first term corresponds to the case of a flat potential 
\cite{linde}. Hence, we see that, provided \mbox{$m_\phi<H/\sqrt{\Delta N}$}
the vector field obtains a mass
\begin{equation}
m\sim g\sqrt{\Delta N}\,H
\end{equation}
where, if inflation is not very long, one might have 
\mbox{$g\sqrt{\Delta N}\sim{\cal O}(1)$}. However, this mass is not tachyonic
and cannot satisfy the condition in Eq.~(\ref{mH}).

One may envisage a possibility to obtain a tachyonic mass in the context of
supergravity, where the kinetic term of the scalar fields is multiplied by the
K\"{a}hler metric:
\begin{equation}
{\cal L}_{D\phi}=K_{\phi\phi^*}D_\mu\phi(D^\mu\phi)^*\,,
\label{LDK}
\end{equation}
Indeed, if the K\"{a}hler potential includes a term of the form
\begin{equation}
\Delta K=-|\phi|^2
\label{K1}
\end{equation}
then we may indeed obtain a tachyonic mass for the vector field. If this mass
is close to satisfying Eq.~(\ref{mH}) it is possible to generate an 
approximately scale invariant spectrum of perturbations. Slight deviations
from the condition in Eq.~(\ref{mH}) generate a tilt on the spectrum, with
spectral index \mbox{$n_s=1+2\tilde\eta$}, where 
\begin{equation}
\tilde\eta=\frac{1}{3}\frac{\tilde m^2}{H^2}=
\frac{1}{3}\left(2+\frac{m^2}{H^2}\right),
\end{equation}
where we used Eq.~(\ref{mtild}).

However, in view of Eqs.~(\ref{LDK}) and (\ref{K1}), we see that a negative 
mass-squared for the vector field requires a negative kinetic term for the 
scalar field $\phi$. Hence, $\phi$ is rendered a ghost field, with all the
unpleasant side effects this might imply (e.g. breaking of Lorentz invariance).
Still, this scenario may find use in the context of the so-called, ghost 
inflation model, which uses exactly such a field \cite{ghost}. However, it is
not clear whether a ghost condensate would still satisfy Eq.~(\ref{phi2}).

Even if we manage to account for all the above we still need to ensure that
the vector field dominates the Universe after the end of inflation, without
causing a large-scale anisotropy. However, having a tachyonic mass as 
described above does not lead to the oscillating behaviour which we would
prefer after the end of inflation. 
To remedy this, we can add another term in the K\"{a}hler potential, coupling
the Higgs field to some other scalar $\psi$:
\begin{equation}
\Delta K=-|\phi|^2+\frac{|\phi|^2|\psi|^2}{M^2},
\end{equation}
where $M$ is an appropriate cutoff scale.
We can then assume that the scalar field $\psi$ is heavy during inflation
and, therefore, \mbox{$\psi\simeq 0$}. However, at or after the end of 
inflation a phase transition gives non-zero vacuum expectation values
(VEVs) to both scalar fields. 
In this scenario the vacuum mass of the vector field would be
\begin{equation}
m^2=g^2\left(\frac{M^2_\psi}{M^2}-1\right)M_\phi^2,
\end{equation}
where $M_\phi$ and $M_\psi$ are the VEVs of $\phi$ and $\psi$ respectively.
From the above we see that there is a chance that, after inflation, the mass 
of the vector field ceases to be tachyonic provided \mbox{$M_\psi\gsim M$}.
This also means that, in the vacuum, $\phi$ is not a ghost.

A similar way to attain the desired results but without casting doubt in the
validity of Eq.~(\ref{phi2}) is as follows. One can consider that instead of
the mass term changing signs after inflation, the kinetic term of the vector 
field does so. Indeed, in supergravity the kinetic term of a vector field
is multiplied by the gauge kinetic function $f$:
\begin{equation}
{\cal L}_{\rm kin}=-\frac{1}{4}fF_{\rm \mu\nu}F^{\mu\nu}.
\label{Lkin}
\end{equation}
The gauge kinetic function is a holomorphic function of the fields of the 
theory. Suppose that it is of the form
\begin{equation}
f(\psi)=-1+\frac{\psi^2}{M^2}\,,
\label{f}
\end{equation}
where $M$ is an appropriate cutoff scale. Then, similarly as before, we may 
assume that $\psi$ is heavy during inflation and, therefore, driven to zero. 
This introduces a change of sign of the kinetic term of the vector field. 
Suppose, now, that the mass of the vector field is \mbox{$m\approx\sqrt 2H_*$},
where $H_*$ is the Hubble scale during inflation. Then, for our considerations,
the reversal of the sign of the kinetic term is entirely equivalent to the 
reversal of the sign of the mass-term because changing the overall sign of the 
vector Lagrangian density in 
Eq.~(\ref{L}) does not affect the equations of motion in Eq.~(\ref{FE}).
Therefore, particle production of the vector field will take place during 
inflation. However, the physics will be affected because one has to consider 
that the sign for the gravitational Lagrangian density is not reversed and we 
may still run into trouble in the same manner as with the ghost condensate 
above. Now, after inflation, a phase transition may send $\psi$ to a non-zero
value $M_\psi$. This changes the sign of the kinetic term 
(renderring the vacuum stable) provided \mbox{$M_\psi\gsim M$}. 
In this case, in the post-inflation era our vector 
field undergoes the desired oscillations and can become a successful curvaton.

Another way to comply with the requirement in Eq.~(\ref{mH}) is to 
assign a ``potential'' $U(\xi)$ to the vector field, in the manner considered 
in Refs.~\cite{VI,Lidsey}, where \mbox{$\xi\equiv A_\mu A^\mu$}. For example,
that way we may assume a negative mass-squared with the vacuum stabilised 
through a self-coupling of the vector field of the form:
\begin{equation}
{\cal L}=-\frac{1}{4}F_{\rm \mu\nu}F^{\mu\nu}-\frac{1}{2}m^2A_\mu A^\mu
+\frac{1}{4}\lambda(A_\mu A^\mu)^2,
\end{equation}
i.e. we assume \mbox{$U=-\frac{1}{2}m^2\xi+\frac{1}{4}\lambda\xi^2$},
with \mbox{$m\approx\sqrt 2H_*$}. Such self-couplings might arise when 
considering non-Abelian vector fields. Note that, the above assumption 
generates a positive mass-squared in the vacuum for our vector field. 
This means that, while the vector field may lie on top of the $U$--potential 
hill during inflation, after inflation $H(t)<m$ and the vector field
begins oscillating in a potential \mbox{$U\propto\xi$}, corresponding to 
isotropic pressureless matter as we have shown.\footnote{If the vector field 
is oscillating into a $U$-potential not linear to $\xi$ then one expects 
%that the average pressure $\overline{p_\perp}$ is not zero.
$\overline{p_\perp}\neq 0$.
Due to 
%the difference of the sign of the pressure in the longitudinal and
%the transverse directions [c.f. Eq.~(\ref{Tdiag})] 
Eq.~(\ref{Tdiag})
%a non-zero $\overline p_\perp$ 
this is bound to generate a large-scale anisotropy if the 
oscillating vector field dominates the Universe. It is intriguing that a weak
such anisotropy (preferred axis) is claimed to exist in the CMB radiation
\cite{joao}.}

Finally, note that the vector field may not need to 
oscillate and dominate the Universe after inflation, in order to imprint
its curvature perturbation spectrum. One may consider a modulated-reheating
scenario \cite{modreh}, where the vector field controls the decay rate of the 
inflaton through a coupling such as the one in Eq.~(\ref{Lm}).

It is evident that the above scenarios are quite contrived and 
involve a number of tunings. 

\section{Conclusions}

In summary we have shown that, in principle, a vector field can indeed be
responsible for the observed curvature perturbation in the Universe. In order
to be so the mass-squared of the vector field has to be 
\mbox{$m^2\approx -2H^2$} during inflation. The vector field must be 
subdominant during inflation to avoid generating a large-scale anisotropy.
Hence, particle production generates an originally isocurvature perturbation,
which can become adiabatic if, after inflation, the vector field dominates the 
Universe before its decay, in accordance to the curvaton scenario. Indeed, we
have shown that, if after inflation the mass-squared of the vector field 
becomes positive, then the field begins oscillating. We have demonstrated that 
an oscillating vector field scales as pressureless matter with the Universe
expansion and does not cause any large scale anisotropy even when it dominates
the Universe. Hence, provided the mass of the vector field complies to the
above requirements, the `vector curvaton' scenario can account successfully for
the observations. 

Admittedly, the condition for the mass of the vector field during inflation is 
hard to achieve and, at best, amounts to a certain level of tunning. Still, 
using a `vector curvaton' may be additionally motivated by the fact that no 
scalar fields have been observed as yet in nature.

\begin{acknowledgements}
I would like to thank D.~H.~Lyth and J.~McDonald for discussions and the 
referee for insightful comments. 
\end{acknowledgements}

\begin{thebiblio}{03}

\bibitem{VI}
L.~H.~Ford,
%``Inflation Driven By A Vector Field,''
Phys.\ Rev.\ D {\bf 40} (1989) 967;
%%CITATION = PHRVA,D40,967;%%
C.~M.~Lewis,
%``Vector inflation and vortices,''
Phys.\ Rev.\ D {\bf 44} (1991) 1661.
%%CITATION = PHRVA,D44,1661;%%

\bibitem{Lidsey}
J.~E.~Lidsey,
%``Cosmological density perturbations from inflationary universes driven by a
%vector field,''
Nucl.\ Phys.\ B {\bf 351} (1991) 695;
%%CITATION = NUPHA,B351,695;%%
A.~B.~Burd and J.~E.~Lidsey,
%``An Analysis of inflationary models driven by vector fields,''
Nucl.\ Phys.\ B {\bf 351} (1991) 679.
%%CITATION = NUPHA,B351,679;%%

\bibitem{triad}
M.~C.~Bento, O.~Bertolami, P.~V.~Moniz, J.~M.~Mourao and P.~M.~Sa,
%``On the cosmology of massive vector fields with SO(3) global symmetry,''
Class.\ Quant.\ Grav.\  {\bf 10} (1993) 285.
%[arXiv:gr-qc/9302034].
%%CITATION = GR-QC 9302034;%%

\bibitem{curv}
D.~H.~Lyth and D.~Wands,
%``Generating the curvature perturbation without an inflaton,''
Phys.\ Lett.\ B {\bf 524} (2002) 5;
K.~Enqvist and M.~S.~Sloth,
%``Adiabatic CMB perturbations in pre big bang string cosmology,''
Nucl.\ Phys.\ B {\bf 626} (2002) 395;
%[arXiv:hep-ph/0109214].
%%CITATION = HEP-PH 0109214;%%
T.~Moroi and T.~Takahashi,
%``Effects of cosmological moduli fields on cosmic microwave background,''
Phys.\ Lett.\ B {\bf 522} (2001) 215
[Erratum-ibid.\ B {\bf 539} (2002) 303];
%[arXiv:hep-ph/0110096].
%%CITATION = HEP-PH 0110096;%%
S.~Mollerach,
%``Isocurvature Baryon Perturbations And Inflation,''
Phys.\ Rev.\ D {\bf 42} (1990) 313.
%%CITATION = PHRVA,D42,313;%%

\bibitem{BD}
T.~S.~Bunch and P.~C.~W.~Davies,
%``Quantum Field Theory In De Sitter Space: Renormalization By Point
%Splitting,''
Proc.\ Roy.\ Soc.\ Lond.\ A {\bf 360} (1978) 117.
%%CITATION = PRSLA,A360,117;%%

\bibitem{linde}
A.~D.~Linde,
%``Scalar Field Fluctuations In Expanding Universe And The New Inflationary
%Universe Scenario,''
Phys.\ Lett.\ B {\bf 116} (1982) 335.
%%CITATION = PHLTA,B116,335;%%

\bibitem{ghost}
N.~Arkani-Hamed, H.~C.~Cheng, M.~A.~Luty and S.~Mukohyama,
%``Ghost condensation and a consistent infrared modification of gravity,''
JHEP {\bf 0405} (2004) 074;
%[arXiv:hep-th/0312099].
%%CITATION = HEP-TH 0312099;%%
N.~Arkani-Hamed, P.~Creminelli, S.~Mukohyama and M.~Zaldarriaga,
%``Ghost inflation,''
JCAP {\bf 0404} (2004) 001.
%%[arXiv:hep-th/0312100].
%%%CITATION = HEP-TH 0312100;%%
%L.~Senatore,
%%``Tilted ghost inflation,''
%Phys.\ Rev.\ D {\bf 71} (2005) 043512.
%%[arXiv:astro-ph/0406187].
%%%CITATION = ASTRO-PH 0406187;%%

\bibitem{modreh}
G.~Dvali, A.~Gruzinov and M.~Zaldarriaga,
%``Cosmological perturbations from inhomogeneous reheating, freezeout, and  
%mass domination,''
Phys.\ Rev.\ D {\bf 69} (2004) 083505;
%[arXiv:astro-ph/0305548].
%%CITATION = ASTRO-PH 0305548;%%
G.~Dvali, A.~Gruzinov and M.~Zaldarriaga,
%``A new mechanism for generating density perturbations from inflation,''
Phys.\ Rev.\ D {\bf 69} (2004) 023505;
%[arXiv:astro-ph/0303591].
%%CITATION = ASTRO-PH 0303591;%%
L.~Kofman,
%``Probing string theory with modulated cosmological fluctuations,''
%arXiv:
astro-ph/0303614;
%%CITATION = ASTRO-PH 0303614;%%
K.~Enqvist, A.~Mazumdar and M.~Postma,
%``Challenges in generating density perturbations from a fluctuating  inflaton
%coupling,''
Phys.\ Rev.\ D {\bf 67} (2003) 121303.
%[arXiv:astro-ph/0304187].
%%CITATION = ASTRO-PH 0304187;%%

\bibitem{joao}
K.~Land and J.~Magueijo,
%``The axis of evil,''
Phys.\ Rev.\ Lett.\  {\bf 95} (2005) 071301.
%[arXiv:astro-ph/0502237].
%%CITATION = ASTRO-PH 0502237;%%

\end{thebiblio}
\end{document}